\documentclass[twocolumn,showpacs,preprintnumbers,amsmath,amssymb]{revtex4}

\usepackage{graphicx}
\usepackage{dcolumn}
\usepackage{bm}

\begin{document}

\preprint{APS/123-QED}

\title{Experimental demonstration of counterfactual quantum key distribution}
\author{Min Ren}
\affiliation{State Key Laboratory of Precision Spectroscopy, East
China Normal University, Shanghai 200062, China}
\author{Guang Wu}
\email{gwu@phy.ecnu.edu.cn}
\affiliation{State Key Laboratory of
Precision Spectroscopy, East China Normal University, Shanghai
200062, China}
\author{E Wu}
\affiliation{State Key Laboratory of Precision Spectroscopy, East
China Normal University, Shanghai 200062, China}
\author{Heping Zeng}
\email{hpzeng@phy.ecnu.edu.cn} \affiliation{State Key Laboratory
of Precision Spectroscopy, East China Normal University, Shanghai
200062, China}

\date{\today}
\begin{abstract}
Counterfactual quantum key distribution provides natural advantage
against the eavesdropping on the actual signal particles. It can
prevent the photon-number-splitting attack when a weak coherent
light source is used for the practical implementation. We realized
the counterfactual quantum key distribution in an unbalanced
Mach-Zehnder interferometer of 12.5-km-long quantum channel with a
high-fringe visibility of $96.4\%$. As a result, we obtained
secure keys against the noise-induced attack (eg. the vacuum
attack) and passive photon-number-splitting attack.
\end{abstract}
\pacs{03.67.Dd }

\maketitle  \noindent

Quantum key distribution (QKD) provides an unconditionally secure
communication between two remote parties (Alice and Bob), where
the security is guaranteed by the fundamentals of quantum
mechanics \cite{Bennett_84,Gisin_02}. However, current techniques
cannot support to implement the ideal quantum cryptography
experiment as originally proposed due to the lack of efficient
single-photon or entangled photon-pair sources in the
near-infrared region. So far, in practical long-distance
fiber-based QKD systems, weak coherent light sources have been
used instead of ideal single photons
\cite{Gobby_04,Mo_05,Peng_07}. Intelligent methods have been
invented to prevent the photon-number-splitting (PNS) attack based
on the multi-photon pulse of the weak coherent light source
\cite{Acin_04,Koashi_04,Guerreau_05,Wu_06}. Among them, the
decoy-state QKD protocol \cite{Hwang_03,Wang_05,Lo_05} could
prevent the PNS attack by statistical security analysis of a large
yield of the photon clicks \cite{Ma_05,Wang_052}. On the other
hand, quantum counterfactual effect has been discussed in
interaction-free measurement \cite{Elitzur_93,Kwiat_95,Penrose_94}
and developed for the quantum computation
\cite{Jozsa_99,Mitchison_01,Hosten_06}. Recently, an interesting
counterfactual QKD protocol was proposed \cite{Noh_09} to
distribute secret keys without any secret information-carrier
qubits transmitting in the quantum channel, which can in principle
prevent the eavesdropper to directly access the entire quantum
system of each qubit. For example, secret key can be established
by using the quantum counterfactual effect in a Michelson
interferometer between Alice and Bob. One arm of the
interferometer is in Alice's secure site, and the other arm is
used to connect Bob as the quantum channel. Unlike previous QKD
protocols which generate secret keys by transmitting the signal
particles through the actual quantum channel, the proposed
counterfactual QKD generates secret keys in the case that the
photonic qubits only transmit in Alice's arm but never travel
through the quantum channel to Bob. Interestingly, as a weak
coherent light source is used, the quantum counterfactual effect
provides a natural advantage to prevent the PNS attack by avoiding
Eve's access to the information-carrier photons. Experimental
demonstration of such a counterfactual QKD requires a stabilized
Michelson interferometer \cite{cho_09}.

In this letter, we experimentally demonstrate the counterfactual QKD
based on a round-way unbalanced Mach-Zehnder interferometer with
$25$~km fiber length difference between the long and short arms that
ensured an effective 12.5-km-long quantum channel between Alice and
Bob. Possible eavesdropping against the counterfactual QKD was
analyzed. The counterfactual QKD scheme was implemented with the
capability to reveal the vacuum attack by monitoring the photon
detection distributions. With a high-fringe visibility of $96.4\%$,
we could obtain secure keys against the passive PNS attack by an
average photon number of $1$ per pulse.

\begin{figure}[b]
\begin{center}
\includegraphics[width=0.45\textwidth]{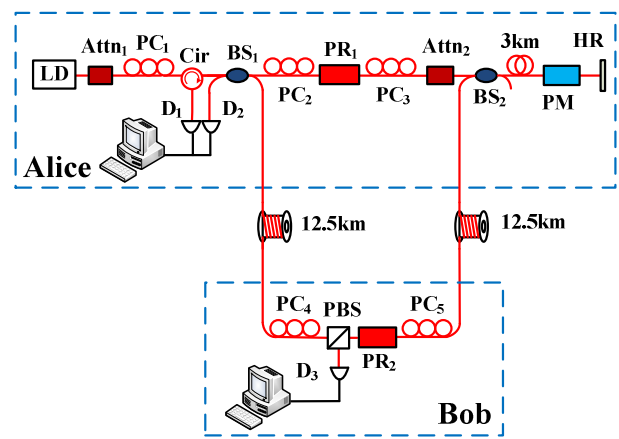}
\caption{Schematic of the counterfactual QKD system. LD: 1550-nm DFB
laser diode; Attn$_{1,2}$: variable attenuators; PC$_{1-5}$: manual
polarization controllers; Cir: circulator; BS$_{1,2}$: 50:50 beam
splitters; PR$_{1,2}$: polarization rotators; PM: phase modulator;
HR: high-reflection mirror; PBS: polarization beam splitter;
D$_{1-3}$: single-photon detectors.} \label{fig_1}
\end{center}
\end{figure}

The counterfactual QKD system was composed of a round-way
unbalanced Mach-Zehnder interferometer as shown in
Fig.~\ref{fig_1}. A 1550-nm DFB laser diode (LD) generated a
series of short pulses with the repetition rate of $5$~kHz, which
were adjusted to horizontal polarization at the input of the
interferometer. The laser pulses were reflected by a
high-reflection mirror (HR), and attenuated to single-photon level
before returning the interferometer. A pulse from the laser source
might travel one of the four paths by the selection of the short
arm ($s$) or long arm ($l$) in the forward and backward
transmissions, denoted as ($ss$), ($sl$), ($ls$), and ($ll$),
respectively. We disregarded the photon paths $ll$ and $ss$ in the
counterfactual QKD, since $ll$ states contained no information,
and $ss$ states were in Alice's secure station with no information
leak. Only the photons travelling $sl$ and $ls$ paths contributed
to QKD, and the interference was automatically stabilized since
photons travelled the fiber paths of exactly the same length. Four
manual polarization controllers (PC$_{2-5}$) were adjusted for
compensation of the polarization drifts in the long-distance fiber
to make the photons of $sl$ and $ls$ paths interfere with the same
polarizations at the output of the static interferometer.

The static interferometer was then used to polarization-encode the
qubit in both Alice's and Bob's sites with two polarization rotators
(PR$_1$ and PR$_2$), which were randomly controlled to rotate the
polarization state by either $0^\circ$ or $90^\circ$. In the forward
transmission, Alice randomly modulated the laser pulses passing the
short arm to the horizontal or vertical polarization direction. Then
two possible quantum states of the photon pulses were prepared as
\begin{equation}
\begin{aligned}
|\psi_0\rangle=(|0\rangle_{sl}|H\rangle_{ls}+|H\rangle_{sl}|0\rangle_{ls})/\sqrt{2},\\
|\psi_1\rangle=(|0\rangle_{sl}|H\rangle_{ls}+|V\rangle_{sl}|0\rangle_{ls})/\sqrt{2},\\
\end{aligned}
\label{Eq_1}
\end{equation}
where $|0\rangle$ is the vacuum state, $|H\rangle$ and $|V\rangle$
denote the horizontal and vertical polarization states, defined as
the bit value of $0$ and $1$, respectively. In the backward
transmission, Bob decoded the polarization states with the
polarization rotator ($PR_2$) and the polarization beam splitter
(PBS) by randomly applying a $90^\circ$ or $0^\circ$ polarization
rotation on the split pulse in the long arm, corresponding to the
bit value of $0$ or $1$. If Alice and Bob chose the same bit
value, the split pulse in the long arm was reflected to the
single-photon detector D$_3$ by the polarization beam splitter.
The quantum states $|\psi_0\rangle$ and $|\psi_1\rangle$ both
collapsed to $|V\rangle_{sl}|0\rangle_{ls}$ or
$|0\rangle_{sl}|H\rangle_{ls}$. The photon travelled either the
short arm, which was then detected by the single-photon detector
D$_1$ or D$_2$ with the same probability of $1/4$, or the long
arm, which was then detected by the single-photon detector D$_3$
with the probability of $1/2$. If Alice and Bob chose different
bit values, the split pulse in the long arm passed through the
polarization beam splitter. The quantum state kept the
interference, and the photon pulses went toward the single-photon
detector D$_1$, wherein $|\psi_0\rangle$ was unchanged, and
$|\psi_1\rangle$ was transformed to $|\psi_0\rangle$ since the
polarization of the split pulse in the long arm was rotated by
$90^\circ$.

Only the events that D$_2$ alone detected a photon would create
the sifted keys, indicating that Alice and Bob have certainly
chosen the same bit value and the photon pulses travelled the
short arm. If D$_1$ or D$_3$ clicked, Alice and Bob announced the
detected results and their bit values. The performance of the
interference could be calculated from these events to monitor
whether there were any Eve's disturbance. We defined ($C1$, $C3$,
$C5$) and ($C2$, $C4$, $C6$) as the counts of the detectors
(D$_1$, D$_3$, D$_2$) in the case that Alice and Bob chose the
same and different bit values, respectively. The count of D$_2$
was randomly selected from the half of the events that D$_2$
clicked, and the other half events created the sifted keys. The
performance of the interference and polarization extinction ratio
of Alice's and Bob's polarization operation could be respectively
characterized by $C1:C2$ and $C3:C4$, where $C1:C2=1:4$ represents
an ideal interference \cite{Calcu1}). The interference fringe
visibility could be calculated by $(4C5-C6)/(4C5+C6)$, while the
error rate of the QKD system could be calculated by $C5/(C5+C6)$.

As Eve had no access to the short arm and the information-carrier
photons, many possible attacks that threaten the conventional QKD,
such as intercept-resent and Trojan horse attacks, could be
successfully prevented \cite{Noh_ar}. Eve may implement a possible
eavesdropping by using ``vacuum attack", in which Eve operates the
polarization-encoding on the vacuum state before Bob's site. Eve's
choice of bit value ($B_{Eve}$) may be the same or different with
Alice's ($B_{Alice}$) or Bob's ($B_{Bob}$). There are altogether
four possible cases. (i) $B_{Eve}=B_{Alice}=B_{Bob}$, Eve may
eavesdrop the sifted key with no errors for Alice and Bob. (ii)
$B_{Eve}=B_{Alice}\neq B_{Bob}$, the quantum state collapses,
while Eve may eavesdrop the sifted key and induce an error for
Bob. (iii) $B_{Eve}\neq B_{Alice}=B_{Bob}$, the photon keeps the
interference and is detected by D$_1$. (iv) $B_{Eve}\neq
B_{Alice}\neq B_{Bob}$, the quantum state collapses, while Eve may
eavesdrop a sifted key with a possible error. Three of the four
cases may create a sifted key, wherein Bob and Eve both have the
error rate of $1/3$ in the sifted key. According to the Shannon
information theory, the information Eve and Bob get is described
as
\begin{equation}
I=1+Dlog_2(D)+(1-D)log_2(1-D), \label{Eq_2}
\end{equation}
where $D$ is the error rate. Here Eve's information through the
vacuum attack is only $8.17\%$, which is much less than $50\%$ in
the conventional QKD through the intercept-resent attack.
Moreover, the vacuum attack changes the count of D$_1$ and D$_2$,
as well as $C1:C2 \longrightarrow 1:2$ \cite{Calcu1}.

The counterfactual cryptography naturally prevents the PNS attack
when a weak coherent light source is used, because Eve even cannot
detect the photon number of each pulse in the long arm, and she has
no access to the short arm \cite{Noh_09}. Passive PNS attack would
provide Eve some information by inserting a beam splitter in the
quantum channel to passively split the photon and replacing the
quantum channel with her lossless channel to compensate the
splitting loss. Eve's gain of the sifted key ($P_{Eve}$) is
extremely limited because only the joint event that there is at
least one photon in each interferometer arm is useful in this
attack. The proportion of Eve's information to the total sifted key
is given by
\begin{equation}
P_{Eve}\leq P(\mu_{sl},n>0) \cdot \eta_{loss},
 \label{Eq_3}
\end{equation}
where $\mu_{sl}$ is the mean photon number of the photon pulse in
the long arm, $\mu$ is the mean photon number of $|\psi_0\rangle$
or $|\psi_1\rangle$. Here $\mu_{sl}=0.5\cdot\mu$, if the long and
short arms have the same attenuation. $P(\mu, n)=e^{-\mu}\mu^n/n!$
is the Poissonian distribution for the photon number of $n$.
$\eta_{loss}=1-10^{-0.25L_{eff}/10}$ is the transmission loss of
the quantum channel from BS$_2$ to PC$_3$, where $L_{eff}$ refers
to the effective distance between Alice and Bob. When the loss of
the quantum channel is so large that $\eta_{loss}\rightarrow 1$,
Eve could nearly obtain all the information of the photon in the
long arm, and her information gain won't increase with the
communication distance any more.

\begin{figure}[b]
\begin{center}
\includegraphics[width=0.4\textwidth]{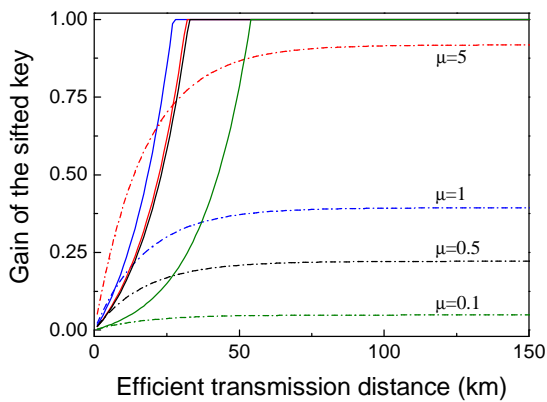}
\caption{Eve's gain of the sifted key. Dashed and solid lines show
Eve's gain through passive PNS attack against the counterfactual
and PNS attack against conventional QKD, respectively. The olive,
black, blue, and red lines denote the mean photon number of $0.1$,
$0.5$, $1.0$, and $5.0$, respectively.} \label{fig_2}
\end{center}
\end{figure}

According to the Shannon information theory, the secure key can be
distilled if Bob obtains more information than Eve. As shown in
Fig.~\ref{fig_2}, Eve's maximum gain of the sifted key was less
than $5\%$ for the case $\mu=0.1$, independent on the large loss
of the long-distance channel. As a result, the counterfactual QKD
protocol provides an approach to the long-distance secure
communication based on the quantum property instead of the
statistical method such as decoy-state protocols.

\begin{figure}
\begin{center}
\includegraphics[width=0.4\textwidth]{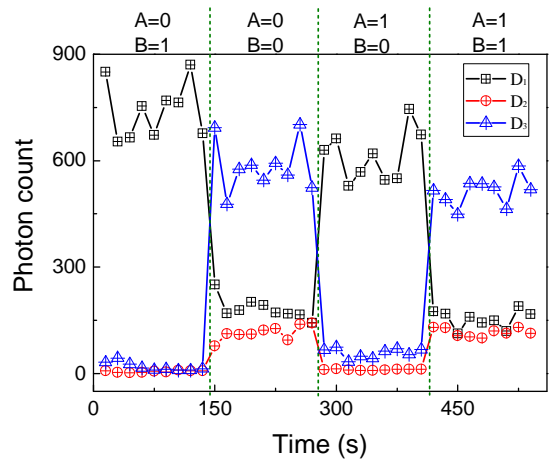}
\caption{Photon counts of D$_1$, D$_2$ and D$_3$ at various bit
value choice of Alice and Bob with the acquisition time of $15$~s.}
\label{fig_3}
\end{center}
\end{figure}

In the experiment, the total loss of the long arm was about
$9$~dB, while the loss of the short arm was only $3$~dB. The
photon flux in the long arm was larger than that in the short arm
in the backward transmission, resulting in an increase of Eve's
gain. To eliminate the threaten induced by the unbalance of the
two arms, we inserted an attenuator (Attn$_2$), and adjusted the
total loss of the short arm to be the same as that of the long
arm. Three home-made near-infrared single-photon detectors were
used with the dark-count probabilities of $1.0\times10^{-5}$,
$5\times10^{-6}$, $1.5\times10^{-5}$ per pulse, respectively, at a
detection efficiency of $10\%$. Due to the birefringence of the
single-mode fiber, the photon pulses of $ls$ or $sl$ path would
cover a different phase shift when Alice sent the horizontal or
vertical polarization state. As there was about $0.4\pi$ phase
shift between two orthogonal polarization in $12.5$~km single-mode
fiber, we used a phase modulator to compensate the
polarization-dependent phase shift. The laser pulses were
attenuated to contain $1$ photon per pulse on average. Then we
measured the single-photon interference and the polarization
extinction ratio of the quantum system. As shown in
Fig.~\ref{fig_3}, when Alice and Bob chose the same or different
polarization, the photon pulses in the long arm were switched to
D$_3$, or passed through the polarization beam splitter to the
output of the interferometer. The photon count of D$_3$ showed a
polarization extinction ratio as high as $30:1$. The single photon
interfered at the output of the interferometer when Alice and Bob
chose different polarization. Owing to the passive phase shift
compensation of the round-way Mach-Zehnder interferometer, the
phase shift caused by the slow variation such as random
temperature and stress drifts were auto-compensated. In this way,
we got a stable interference with the fringe visibility of
$96.4\%$. When Alice and Bob chose the same polarization, the
quantum state of the signal photons collapsed, and we got almost
the same count rates on D$_1$ and D$_2$.

The experimental demonstration of the counterfactual QKD was then
carried out with the average photon number $\mu=1.0$. Alice and
Bob used two random number generators to drive PR$_1$ and PR$_2$
independently. Table~\ref{tab_1} presents the experimental results
with an acquisition time of $540$~s. The whole system showed an
unsurpassed stability during the acquisition. The counts of each
single-photon detector and the coincidence counts between D$_3$
and D$_1$ or D$_2$ showed that the photon distribution experienced
no observable changes. The error rate D$_{AB}$ was $6.7\%$, mainly
from the noise of the interference. Note that the counterfactual
QKD requires a single-photon interference of a high-fringe
visibility, which would have fourfold errors as the conventional
QKD of the same fringe visibility. The interference errors were
mainly induced by the polarization operation. The polarization
extinction ratio was kept at $30.5:1$ in the experiment. This
error rate opened the back door for Eve to implement the vacuum
attack against $20\%$ of the vacuum states. However, it would
change the counts of D$_1$ and D$_2$, and $C1:C2 \longrightarrow
1:3.3$ \cite{Calcu1}. In the experiment, we got $C1:C2=4.1$,
implying that the error rate didn't come from the vacuum attack
but was induced by the intrinsic noise of the QKD system. In this
case, Eve may obtain $20\%$ of the information through the passive
PNS attack, while Bob's information was $64.5\%$ according to
Eq.~\ref{Eq_2}. Note that Eve's information was restricted under a
certain level under this attack (Fig.~\ref{fig_2}) and was
insensitive to the transmission loss. Since Bob's information was
much more than Eve's, Alice and Bob could distill the
unconditional secure keys at last by the classical error
correction and privacy amplification methods
\cite{Gisin_02,Bennett_88,Brassard_94}.

\begin{table}
\begin{center}
\caption {Experimental results of $12.5$-km counterfactual QKD
with the acquisition time of $540$~s.}
\begin{tabular}{m{5cm}m{2cm}}

  \hline
  \hline
   D$_1$&15149\\
   D$_2$&2243\\
   D$_3$&10577\\
\hline
   $C1:C2$&1 : 4.1\\
   Fringe Visibility &$96.4\%$\\
   Polarization Extinction Ratio&30.5 : 1\\
   Coincidence Probability&0.15$\%$\\
   Sifted key&1121\\
   $D_{AB}$&6.7$\%$\\
  \hline
  \hline
\end{tabular}\label{tab_1}

\end{center}
\end{table}

In conclusion, we realized the counterfactual QKD experiment in a
round-way unbalanced Mach-Zehnder interferometer of $25$~km fiber
length difference between the long and short arms that ensured a
fiber-based quantum channel of $12.5$~km. Despite Bob had an error
rate of $6.7\%$, we could ensure that there was no noise-induced
attack according to the unchanged count distribution of each
single-photon detectors used in the QKD system, and secure keys
could be obtained against the passive PNS attack. The key
generation rate of the counterfactual QKD, which was mainly
limited by the slow response time of the polarization rotators
used in our experiment, could be increased with fast polarization
rotators and high-speed single-photon detectors
\cite{Yuan_07,Xu_09}. And the active polarization compensating
methods have been invented for long-distance fiber-based QKD
experiments\cite{Chen_07,Xavier_08,Chen_09}, which were quite
useful to realize a long-term stable and long-distance
counterfactual QKD system. The counterfactual QKD scheme was
implemented with currently available technologies, promising a
robust and practical quantum cryptography system toward global
secure communication.

This work was funded in part by National Natural Science Fund of
China (10525416, 10904039 and 10990101), and National Key Project
for Basic Research (2006CB921105).


\begin{thebibliography}{99}
\bibitem{Calcu1}
If there is no eavesdropping, the qubit collapses or interferes with
the same probability (D$_1$ clicks certainly). So
$C1=\frac{1}{2}\cdot\frac{1}{2}\cdot\frac{1}{2}=\frac{1}{8}$, and
$C2=\frac{1}{2}\cdot\frac{1}{2}\cdot1+\frac{1}{2}\cdot\frac{1}{2}\cdot1=\frac{1}{2}$.
If Eve does the vacuum attack, the qubit collapses (or interferes)
with the probability of $\frac{3}{4}$ (or $\frac{1}{4}$). So
$C1=\frac{1}{2}\cdot\frac{3}{4}\cdot\frac{1}{2}=\frac{3}{16}$, and
$C2=\frac{1}{2}\cdot\frac{1}{2}\cdot1+\frac{1}{2}\cdot\frac{1}{4}\cdot1=\frac{3}{8}$.
\end{thebibliography}
\end{document}